\documentclass[twocolumn]{aastex63}
\usepackage{subfigure}
\usepackage{amsmath}
\usepackage{amssymb}

\shorttitle{\tess\, Ephemeris Expiration}
\shortauthors{Dragomir et al.}

\newcommand{\tess}{{\it TESS}}
\newcommand{\jwst}{{\it JWST}}
\newcommand{\kepler}{{\it Kepler}}
\newcommand{\cheops}{{\it CHEOPS}}
\newcommand{\spitzer}{{\it Spitzer}}

\newcommand{\ktwo}{{K2}}
\newcommand{\hst}{{\it HST}}

\begin{document}

\title{Securing the legacy of \tess\, through the care and maintenance of \tess\, planet ephemerides}

\email{dragomir@unm.edu}
\author[0000-0003-2313-467X]{Diana~Dragomir}
\affiliation{Department of Physics and Kavli Institute for Astrophysics and Space Research, Massachusetts Institute of Technology, Cambridge, MA 02139, USA}
\affiliation{Department of Physics and Astronomy, University of New Mexico, Albuquerque, NM, USA}

\author{Mallory~Harris}
\affiliation{New College of Florida, Sarasota, FL, 34243, USA}
\affiliation{Department of Physics and Astronomy, University of New Mexico, Albuquerque, NM, USA}

\author[0000-0002-3827-8417]{Joshua~Pepper}
\affiliation{Department of Physics, Lehigh University, 16 Memorial Drive East, Bethlehem, PA, 18015, USA}

\author[0000-0001-7139-2724]{Thomas~Barclay}
\affiliation{University of Maryland, Baltimore County, 1000 Hilltop Circle, Baltimore, MD 21250, USA}
\affiliation{NASA Goddard Space Flight Center, 8800 Greenbelt Road, Greenbelt, MD 20771, USA}

\author[0000-0001-6213-8804]{Steven~Villanueva,~Jr.}
\affiliation{Department of Physics and Kavli Institute for Astrophysics and Space Research, Massachusetts Institute of Technology, Cambridge, MA 02139, USA}

\author[0000-0003-2058-6662]{George~R.~Ricker}
\affiliation{Department of Physics and Kavli Institute for Astrophysics and Space Research, Massachusetts Institute of Technology, Cambridge, MA 02139, USA}

\author[0000-0001-6763-6562]{Roland~Vanderspek}
\affiliation{Department of Physics and Kavli Institute for Astrophysics and Space Research, Massachusetts Institute of Technology, Cambridge, MA 02139, USA}

\author[0000-0001-9911-7388]{David~W.~Latham}
\affiliation{Center for Astrophysics $|$ Harvard \& Smithsonian, 60 Garden Street, Cambridge, MA 02138, USA}

\author[0000-0002-6892-6948]{S.~Seager}
\affiliation{Department of Physics and Kavli Institute for Astrophysics and Space Research, Massachusetts Institute of Technology, Cambridge, MA 02139, USA}
\affiliation{Department of Earth, Atmospheric and Planetary Sciences, Massachusetts Institute of Technology, Cambridge, MA 02139, USA}
\affiliation{Department of Aeronautics and Astronautics, MIT, 77 Massachusetts Avenue, Cambridge, MA 02139, USA}

\author[0000-0002-4265-047X]{Joshua~N.~Winn}
\affiliation{Department of Astrophysical Sciences, Princeton University, 4 Ivy Lane, Princeton, NJ 08544, USA}

\author{Jon~M.~Jenkins}
\affiliation{NASA Ames Research Center, Moffett Field, CA, 94035, USA}

\author[0000-0002-5741-3047]{David~ R.~Ciardi}
\affiliation{Caltech/IPAC-NASA Exoplanet Science Institute, 770 S. Wilson Avenue, Pasadena, CA 91106, USA}

\author{Gabor~Furesz}
\affiliation{Department of Physics and Kavli Institute for Astrophysics and Space Research, Massachusetts Institute of Technology, Cambridge, MA 02139, USA}

\author{Christopher~ E.~Henze}
\affiliation{NASA Ames Research Center, Moffett Field, CA, 94035, USA}

\author[0000-0002-4510-2268]{Ismael~Mireles}
\affiliation{Department of Physics and Kavli Institute for Astrophysics and Space Research, Massachusetts Institute of Technology, Cambridge, MA 02139, USA}

\author{Edward~H.~Morgan} 
\affiliation{Department of Physics and Kavli Institute for Astrophysics and Space Research, Massachusetts Institute of Technology, Cambridge, MA 02139, USA}

\author{Elisa~Quintana}
\affiliation{NASA Goddard Space Flight Center, 8800 Greenbelt Road, Greenbelt, MD 20771, USA}

\author[0000-0002-8219-9505]{Eric~B.~Ting}
\affiliation{NASA Ames Research Center, Moffett Field, CA, 94035, USA}

\author[0000-0003-4755-584X]{Daniel~Yahalomi}
\affiliation{Center for Astrophysics $|$ Harvard \& Smithsonian, 60 Garden Street, Cambridge, MA 02138, USA}

\begin{abstract}

\noindent Much of the science from the exoplanets detected by the \tess\ mission relies on precisely predicted transit times that are needed for many follow-up characterization studies. We investigate ephemeris deterioration for simulated \tess\ planets and find that the ephemerides of 81\% of those will have expired (i.e. 1$\sigma$ mid-transit time uncertainties greater than 30 minutes) one year after their \tess\ observations. We verify these results using a sample of \tess\ planet candidates as well. In particular, of the simulated planets that would be recommended as \jwst\ targets by \cite{Kem18}, $\sim$80\% will have mid-transit time uncertainties $>$ 30 minutes by the earliest time \jwst\ would observe them. This rapid deterioration is driven primarily by the relatively short time baseline of \tess\ observations. We describe strategies for maintaining \tess\ ephemerides fresh through follow-up transit observations. We find that the longer the baseline between the \tess\ and the follow-up observations, the longer the ephemerides stay fresh, and that 51\% of simulated primary mission \tess\ planets will require space-based observations. The recently-approved extension to the \tess\ mission will rescue the ephemerides of most (though not all) primary mission planets, but the benefits of these new observations can only be reaped two years after the primary mission observations. Moreover, the ephemerides of most primary mission \tess\ planets (as well as those newly discovered during the extended mission) will again have expired by the time future facilities such as the ELTs, Ariel and the possible LUVOIR/OST missions come online, unless maintenance follow-up observations are obtained.

\end{abstract}

\section{Introduction}

Of the nearly 4000 exoplanets known to date, 75\% transit their host star despite the relatively low probability of this favourable alignment. This is largely due to the \kepler\ mission \citep{Borucki}, with help from the Corot mission \citep{Barge} and long-term ground-based transit surveys such as OGLE \citep{Konacki}, SuperWASP \citep{Pollacco}, HATNet/HATSouth \citep{Bakos,Bakos_2013}, KELT \citep{Pep07,Pep12}, MEarth \citep{Nutzman}, TrES \citep{Odonovan} and XO \citep{Mccullough}, as well as the more recent surveys TRAPPIST \citep {Jeh11}, NGTS \citep{Wes16}, and MASCARA \citep{Tal17}.

The \kepler\ sample in particular has greatly advanced our understanding of exoplanet occurrence and system architecture. Major discoveries include evidence that planets smaller than Neptune are more common than larger planets \citep{Fressin13, Petigura13}, the fact that small planets often form in compact multi-planet systems \citep{Lat11, Lis11, Row14}, and the presence of circumbinary planets \citep{Doy11,Welsh}. While immensely significant, these discoveries also raise new questions. To further understand the origins of these planet populations, we need to determine the composition of the planets by measuring their masses, probing their atmospheres, and characterizing their host stars in detail. However, the vast majority of \kepler\ systems are too distant and too faint for these studies.

The recently launched Transiting Exoplanet Survey Satellite (\tess) comes to the rescue with a promise to revolutionize the field of exoplanet research. \tess\, is expected to discover thousands of transiting planets, including several hundred orbiting stars within 100 pc of the Solar System \citep{Sul15,Bar18,Hua18}. Thus, many \tess\, systems are bright and amenable to detailed characterization. In the next few years we will make considerable strides toward a population-level grasp not just of small planets' sizes and period distributions, but also of their masses, atmospheres and their host stars' properties. 

\tess\, is finding transiting planets with a variety of sizes and a relatively wide range of orbital periods, but longer-period transiting planets are more rare due to the reduced probability of transit farther from the host star and finite \tess\ observing baseline. This factor, combined with the desire to study exoplanets across a wide range of equilibrium temperatures, makes the discovery of long-period transiting planets quite valuable. At the same time, given the mission duration and observing strategy, many of the longer-period planets have few transits observed by \tess. All else being equal, long-period planets thus have a greater uncertainty in their periods, as determined from the \tess\, observations alone. This can lead to a larger uncertainty in the mid-transit time after a given stretch of time, relative to a shorter-period planet.

\tess\, planets will be the targets of a variety of follow-up observations, beyond confirmation and mass measurements. Here, we collectively refer to those that depend sensitively on a planet's ephemeris as ``time-sensitive characterization observations" (TCOs). The science goals of TCOs include 
\begin{itemize}
\item atmospheric characterization (particularly through transmission or secondary eclipse spectroscopy)
\item orbital obliquity measurements (through Doppler tomography or the Rossiter-McLaughlin effect)
\item measurements of transit timing and duration (for orbital decay or TTV mass measurements, or searches for exomoons or additional planets in a system)
\item transit parameter refinement (e.g. for improving the precision of the measured planet radius or orbital inclination)
\item characterization of the host star through measurements of limb darkening and starspot properties, as well as constraints on the stellar surface gravity
\end{itemize}

In order to schedule TCOs, particularly those making use of expensive resources like the Hubble Space Telescope (\hst) or the James Webb Space Telescope (\jwst), the mid-transit time should ideally have an uncertainty of less than 30 min. In this paper, we consider a planet's ephemeris to be expired when the 1$\sigma$ uncertainty on its mid-transit time becomes greater than 30 minutes. Such an uncertainty requires devoting an additional 2 hours to any TCOs, in order to have 95\% confidence that the full transit will be observed.

Previous work partly related to the subject of ephemeris deterioration has been published by \cite{Dee17}, who devote a section of their paper to investigating the timing precision of 20 hypothetical 2-minute cadence \tess\, planets observed during one \tess\, pointing (27.4 days), and spanning a range of parameters. Our work differs in several ways. We use the latest planet yield simulations to obtain a bulk picture of the ephemeris deterioration for the entire set of expected \tess\, planets. In so doing, our analysis naturally incorporates the effect of time coverage by multiple 27.4-day sectors, which affects a disproportionate number of simulated \tess\, planets (a selection effect whereby the detectability of a transiting planet increases the longer it is observed). In addition to 2-minute cadence planets, we also examine 30-minute cadence planets, for which ephemeris deterioration is the most severe and the need for rescue is greatest. Finally, while the principal product of \cite{Dee17} is a transit and eclipse timing precision estimator, our aim is to analyze in detail the outcomes of \tess\, ephemeris precision, explore the problem of fast ephemeris deterioration, and propose follow-up strategies for correcting this problem. We also note a white paper by \cite{Bou17} that investigated the impact and yield of various \tess\ extended mission scenarios, and noted that a repeat of the primary mission (PM) would be most beneficial for ephemeris refreshment of primary mission planets.

This paper is organized as follows. In section \ref{sec:TESSsims} we briefly describe the \tess\, mission and the planet yield simulations we used in our analysis. In section \ref{sec:simanalysisresults} we present the details of our analysis and results as a function period, planet size, stellar magnitude and stellar effective temperature, for the simulated planet sample. We also examine the ephemeris deterioration of real \tess\ planet candidates in section \ref{sec:toianalysisresults}. We discuss the implications of those results and the impact of the extended \tess\ mission, and make recommendations for maintaining accurate \tess\, ephemerides in section \ref{sec:discussion}. We summarize our findings and conclude in \ref{sec:conclusion}.

\section{The \tess\, Mission and Yield Simulations}
\label{sec:TESSsims}

\tess\, \citep{Rick15} is a NASA space telescope searching for transiting planets launched in April 2018, with a two-year prime mission.  \tess\, acquires observations in two modes.  A selection of about 200,000 target stars (TS) are observed at a 2-minute cadence, while images of the entire field of view (Full Frame Images, or FFIs) are observed at a 30-minute cadence. The short-cadence target stars are selected as prime targets for transit detection, and are primarily bright and/or cool dwarf stars.

\begin{figure*}[t]
    \centering
    \includegraphics[scale=0.33]{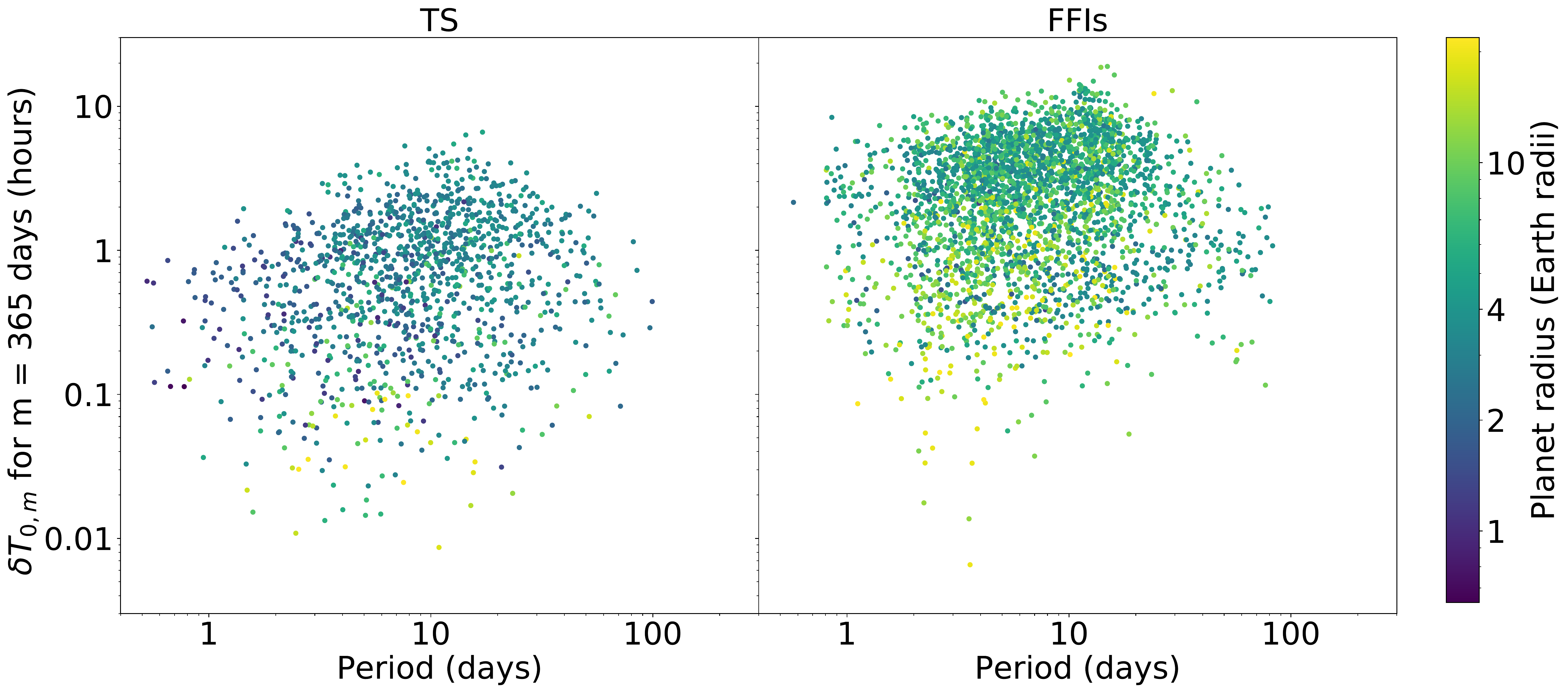}
    \caption{The uncertainty in mid-transit time for simulated planets one year after \tess\, observes them, as a function of orbital period. The colors represent planet radius. {\it Left:} TS planets. {\it Right:} FFI planets. The ephemerides of 61\% of TS and 89\% of FFI planets will have expired one year after their \tess\ observations.}
    \label{fig:ephdecay-summary}
\end{figure*}

\tess\, observes the sky in a set of pointed observations in which the spacecraft nearly continuously observes a section of the sky stretching from 6 degrees from the ecliptic to the ecliptic pole for 27 days, with each section referred to as a sector.  The mission steps around in ecliptic longitude, and has used 13 sectors to cover most of the southern ecliptic hemisphere over the course of a year, and has recently rotated, now observing the northern hemisphere.  Near the ecliptic poles, subsequent sectors overlap, so that stars in those regions can be observed for many months.  The majority of the sky observed by \tess\, (74\%) has an observational time baseline of only $\sim$27 days. For transiting exoplanets with orbital periods longer than 13.5 days seen in only a single sector, \tess\, can only capture one or two transits, and for planets in those regions with periods longer than 27 days, \tess\, can only capture at most one transit. In these cases, the ephemerides of the planets are difficult to determine using \tess\, data alone. 

A number of simulations of the \tess\, planet yield have been carried out: \cite{Sul15}, \cite{Bou17}, \cite{Bar18}, \cite{Mui18}, \cite{Bal19}, \cite{Vil19} and \cite{Hua18}. The simulations from \cite{Bal19} and \cite{Mui18} focused on the planet yield for M dwarfs, while \cite{Vil19} focused on the yield of planets for which only one transit would be observed by \tess, so none of those three yield simulations that are sufficiently general for the scope of this paper. Of the remaining four studies, \cite{Sul15} and \cite{Bou17} drew stars from a Galactic model, while the other two used real stars as listed in the \tess\, Input Catalog \citep[TIC;][]{Sta18} for their simulations. 

Compared to \cite{Bar18}, the simulations of \cite{Hua18} use an updated 2-minute target list, Gaia-updated stellar parameters, more realistic noise parameters and multi-planet system occurrence rates, and stars with \tess\, magnitude as faint as $T=15$.  However, the two works find similar planet yields for bright stars (\citealt{Bar18} only uses stars with \tess\, magnitude brighter than about $T=13$, depending on the stellar temperature). Since we aim to examine statistically how our knowledge of \tess\, planet ephemerides depends on the parameters of the planetary systems, we do not expect our overall results to depend on the number of planets found, only on planetary and stellar parameters.  Since the simulation results of \cite{Hua18} are not currently publicly available while those of \cite{Bar18} are, we select the latter as the basis for our analysis. 

\section{Ephemeris Expiration Analysis and Results Using Planet Yield Simulations}
\label{sec:simanalysisresults}

The simulations of \cite{Bar18} predict that \tess\, will detect 1296 TS planets and 3080 FFI planets (with at least two transits observed by \tess).  

\subsection{Analysis}
For each planet, we determined the signal-to-noise ratio (SNR) for one transit, using the combined SNR (SNRF) and number of \tess\, transits ($N_{transits}$) included in the simulated planet catalog. Next, we calculated ingress duration ($\tau$) using the following formula:
\begin{equation}
\tau = \frac{Rp}{a} \frac{P}{\pi}
\end{equation}
which assumes a circular orbit and an inclination of 90$^{\circ}$ for every planet.
Then, we used the equations of \cite{Pri14} to compute the uncertainty on the mid-transit time ($\delta T_c$) for an individual transit $c$:

\begin{equation}
\label{eqn:sigtc1}
\begin{split}
\delta T_c = \frac{1}{SNR}\sqrt{\frac{\tau T_{dur}}{2}} \frac{1}{\sqrt{1-\frac{I}{3\tau}}}, \tau \geq I \\
\end{split}
\end{equation}

\begin{equation}
\label{eqn:sigtc2}
\begin{split}
\delta T_c = \frac{1}{SNR}\sqrt{\frac{I T_{dur}}{2}} \frac{1}{\sqrt{1-\frac{\tau}{3I}}}, I \geq \tau
\end{split}
\end{equation}
where $T_{dur}$ is the transit duration and $I$ is the integration time. We note that equation \ref{eqn:sigtc1} indicates that for $I$ shorter than $\tau$, $\delta T_c$ decreases with increasing $\tau$. In other words, the better ingress and egress are sampled, the better the precision on the mid-transit time measurement.

\begin{figure*}
    \centering
    \includegraphics[scale=0.37]{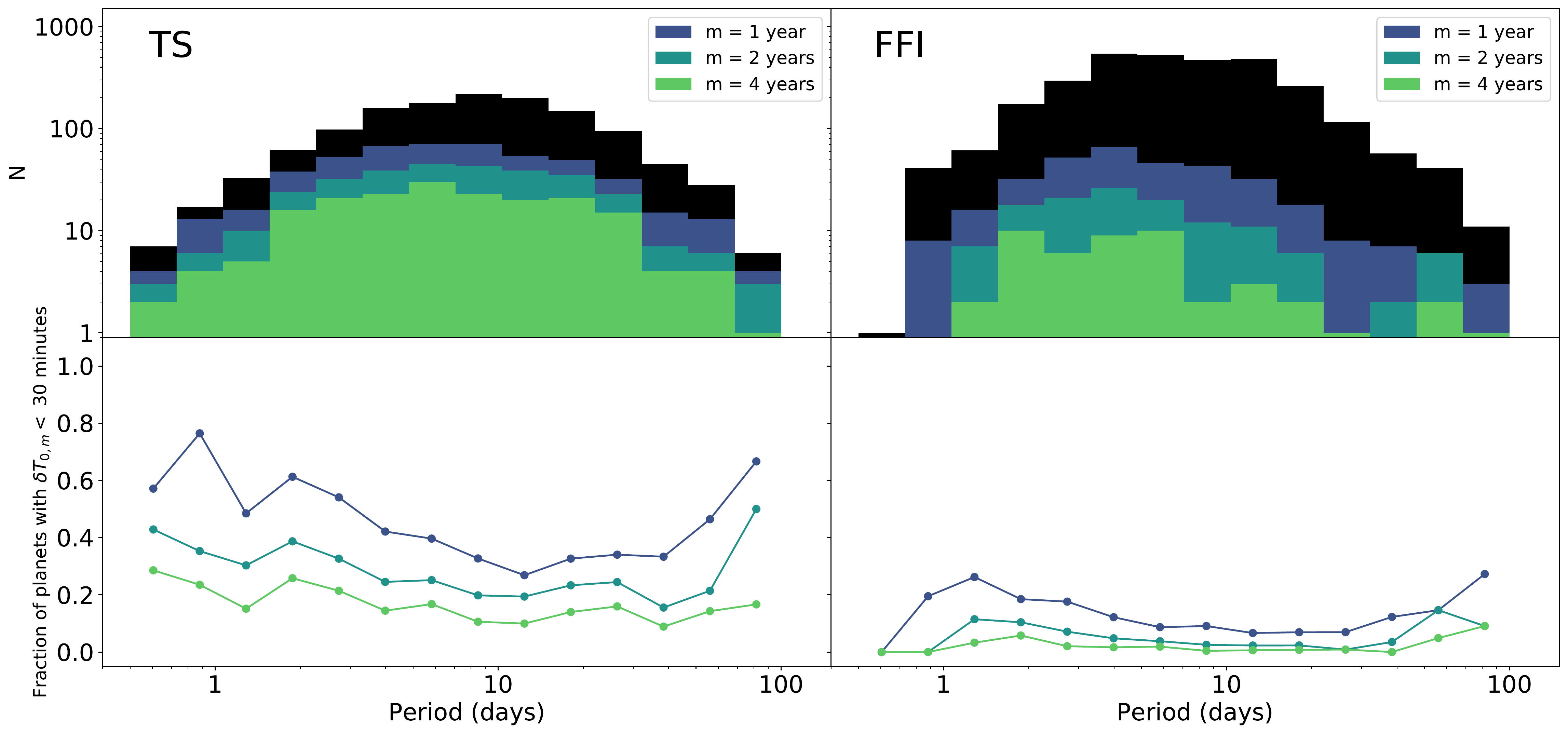}
    \caption{In the top panels, black corresponds to the distribution of the full sample of simulated planets used in our analysis (1296 TS and 3080 FFI planets), as a function of period. Navy, turquoise, and green represent the distributions of planets with $\delta T_{0,m} < $ 30 min. for $m =$ 1, 2 and 4 years (from the end of the \tess\ observations of each planet), as a function of period. The bottom panels show the fraction of planets (with $\delta T_{0,m} < $ 30 min. for $m =$ 1, 2 and 4 years) relative to the full sample, as a function of period. The figures on the left and right correspond to TS and FFI planets, respectively.}
    \label{fig:ephdecay-period}
\end{figure*}

\begin{figure*}
    \centering
    \includegraphics[scale=0.37]{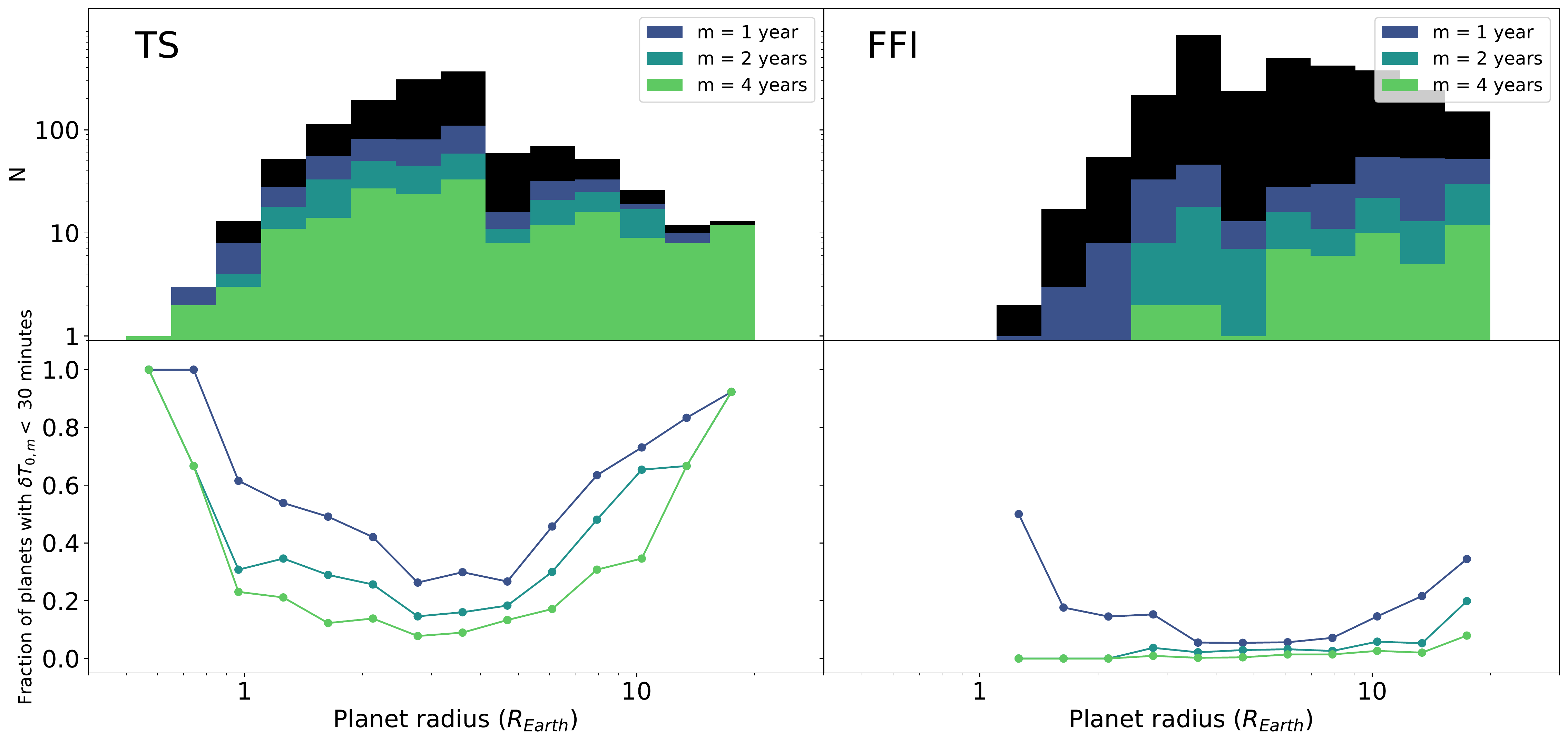}
    \caption{Distribution (top) and fraction (bottom) of planets with $\delta T_{0,m} < $ 30 min, as a function of planet radius for different values of $m$, with TS planets on the left and FFI planets on the right. Colors are as in Figure \ref{fig:ephdecay-period}.}
    \label{fig:ephdecay-radius}
\end{figure*}

\begin{figure*}
    \centering
    \includegraphics[scale=0.37]{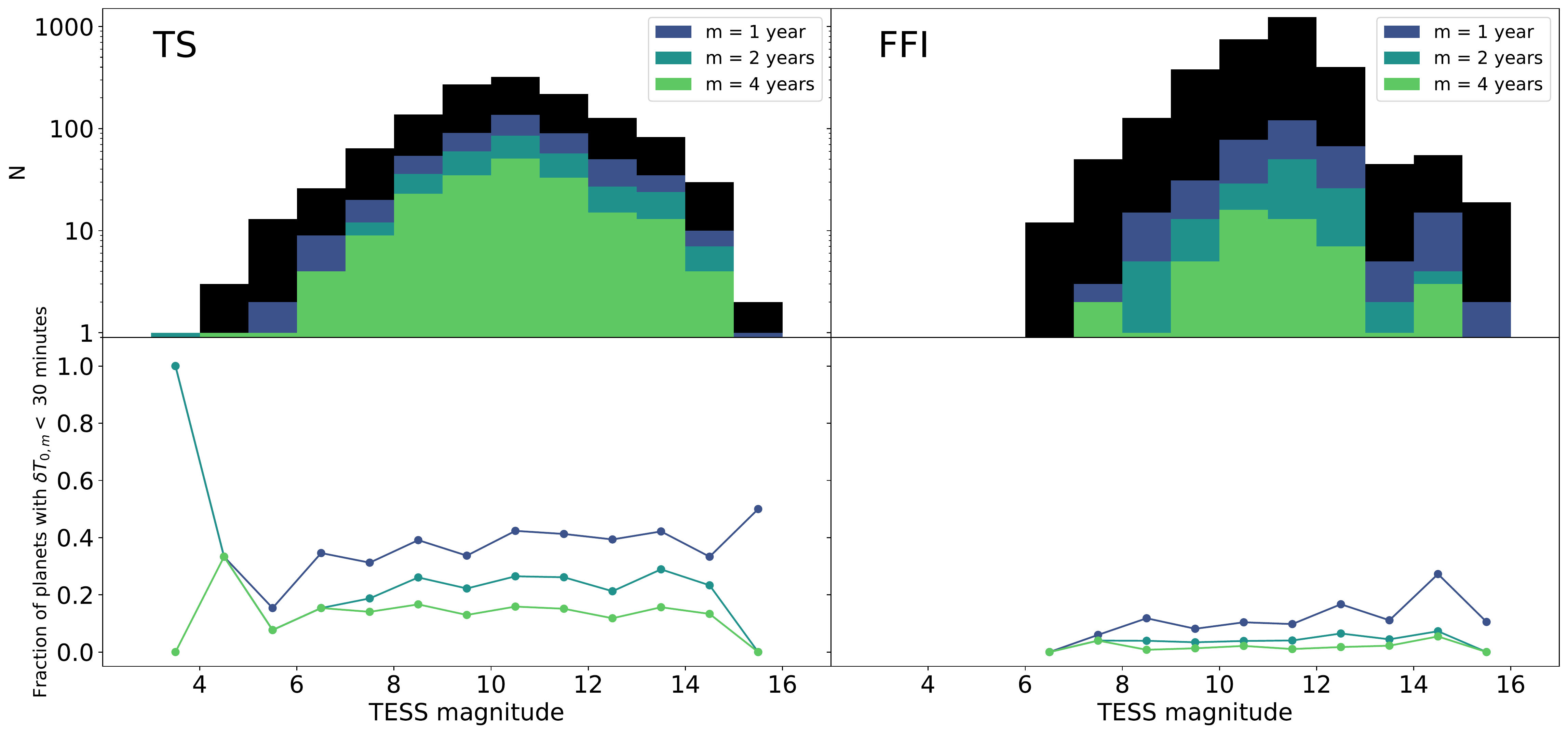}
    \caption{Distribution (top) and fraction (bottom) of planets with $\delta T_{0,m} < $ 30 min, as a function of \tess\, magnitude for different values of $m$, with TS planets on the left and FFI planets on the right. Colors are as in Figure \ref{fig:ephdecay-period}. We remind the reader that our analysis uses the \tess\ planet yield simulations of \cite{Bar18}, who used real stars as listed in the \tess\, Input Catalog (rather than drawing stars from a Galactic model as other works have done) for their simulations (see section 2 for details).}
    \label{fig:ephdecay-tmag}
\end{figure*}

\begin{figure*}
    \centering
    \includegraphics[scale=0.37]{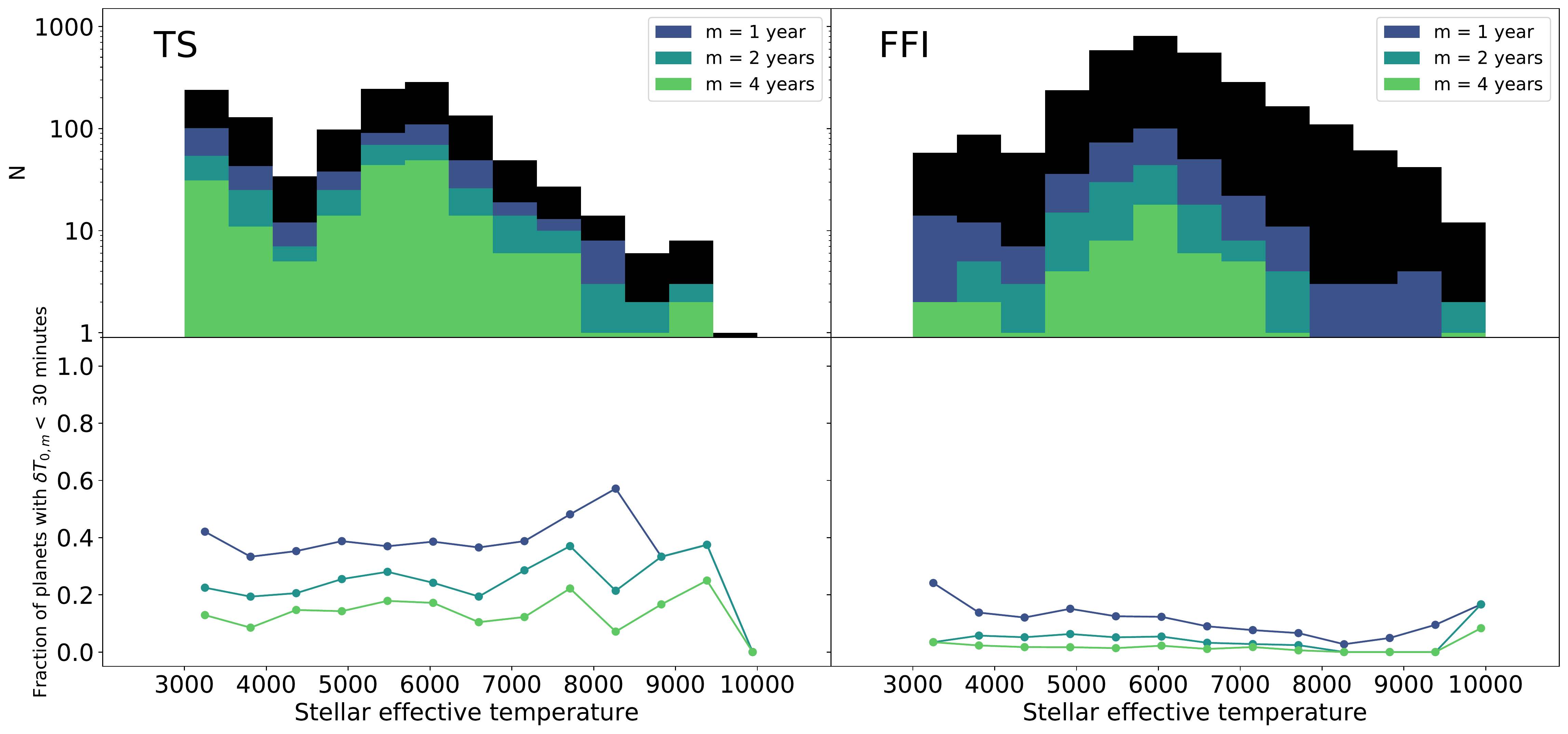}
    \caption{Distribution (top) and fraction (bottom) of planets with $\delta T_{0,m} < $ 30 min, as a function of stellar effective temperature for different values of $m$, with TS planets on the left and FFI planets on the right. Colors are as in Figure \ref{fig:ephdecay-period}.}
    \label{fig:ephdecay-teff}
\end{figure*}

For each planet, at the end of \tess\, observations, there is an ephemeris, represented by the mid-transit time $T_0 \pm \delta T_0$ (corresponding to the epoch of the last \tess\ transit), and associated period $P \pm \delta P$.  This ephemeris is determined from the individual measured times of transit from the \tess\ observations ($T_c \pm \delta T_c$), as follows. All uncertainties ($\delta T_0, \delta P, \sigma_{t_{c}}$) correspond to one standard deviation (1 $\sigma$) from the mean.

For each planet, we generated 1000 sets of $N_{transits}$ mid-transit times as would be observed by \tess. For each simulated transit, we represent the observed mid-transit time as a value drawn from a Gaussian distribution centered on the ``true" mid-transit time and with a standard deviation equal to the calculated $\delta T_c$. We also assign an uncertainty of $\delta T_c$ to each transit. For a given planet, we then fit a linear regression to all transits using least squares minimization. We take the mean of the 1000 best-fit slope values as the best-fit period. We compute the uncertainty on the period ($\delta{P}$) by taking the standard deviation of the distribution of best-fit period values across all 1000 simulations of that planet.

We then determine the uncertainty on a future mid-transit time, $\delta T_{0,m}$, where $m$ represents the time elapsed from the end of the \tess\, observations of a particular target. We calculate $\delta T_{0,m}$ for every simulated \tess\, planet as follows \footnote{$P$ and $T_{0,\tess}$ are not independent variables. However, in our analysis it is not trivial to accurately determine the amount of covariance between them (since we are not analysing simulated light curves, but rather simulated mid-transit times with simulated uncertainties). Therefore, to be conservative and avoid under-estimating the uncertainty on future mid-transit times, we assume $P$ and $T_0$ are fully correlated. We note that the choice of adding linearly vs. in quadrature only changes the future mid-transit time uncertainty by a negligible amount.}: 
\begin{equation}
\label{eqn:delTc}
\delta T_{0,m} = n_{m}\delta P + \delta T_{0,\tess} 
\end{equation}
where $\delta T_{0,\tess}$ is the uncertainty on the mid-point of the last \tess\, transit, $\delta P_{\tess}$ is the uncertainty on the period determined from the \tess\, observations, and $n_{m}$ is the number of planet orbital cycles between $T_{0,\tess}$ and $T_{0,m}$ (see also \citealt{Zel19}). 

Our analysis does not take into account transit timing variations (TTVs) that may occur in multi-planet systems. The amplitude of any TTVs is affected by the masses, periods and orbital eccentricities of planets in the system. However, TTVs are predicted to be uncommon in TESS data \citep{Had19}, and were not incorporated in the \tess\, planet simulations we used here. 

\subsection{Results}

We performed the analysis described above separately for simulated TS and FFI planets. We show $\delta T_{0,m}$ evaluated 1 year after the end of a planet's \tess\, observations ($\delta T_{0,1y}$) as a function of orbital period and planet size (represented by the color gradient) in Figure \ref{fig:ephdecay-summary}, with TS planets on the left and FFI planets on the right.

A few features stand out in Figure \ref{fig:ephdecay-summary}. The higher cadence of the TS observations leads to slower ephemeris deterioration for these planets than for the FFI planets, because $\delta T_c$ is smaller, thus reducing $\delta P$ as well.  We also note that in general, larger planets have smaller $\delta T_{0,1y}$.

The upper range of $\delta T_{0,1y}$ increases with increasing period until just after $P\sim$10 days. For longer periods, $\delta T_{0,1y}$ begins to decrease with period. To explore this further, we examined the fraction of planets with $\delta T_{0,m} < $ 30 min (i.e. the threshold we use to determine whether the ephemeris has deteriorated, as described in section 1), as a function of period, for three different values of $m$ for TS and FFI planets (Figure \ref{fig:ephdecay-period}).  The fraction of planets with $\delta T_{0,m} < $ 30 min decreases as the period increases, but only until $P \sim 10$ days; beyond this threshold, the fraction of planets with $\delta T_{0,m} < $ 30 min increases with period. This trend mirrors the features seen in Figure \ref{fig:ephdecay-summary} and holds for different values of $m$ (though it weakens with increasing $m$). For planets with short periods, a large proportion of candidates have $\delta T_{0,m} < $ 30 min due to TESS observing many transits of these planets, resulting in a smaller initial uncertainty in the period. A large proportion of long-period planets have $\delta T_{0,m} < $ 30 min because, while $\delta P$ may be larger due to TESS observing fewer transits of these planets, these planets also experience fewer orbital cycles during the subsequent time span. We find that the latter effect over-compensates for the former, such that the uncertainty on the future mid-transit time for the longer-period planets does not increase as quickly as it does for planets with intermediate periods.

We also looked at the effects of planet radius (Figure \ref{fig:ephdecay-radius}), host star brightness (Figure \ref{fig:ephdecay-tmag}) and stellar effective temperature (Figure \ref{fig:ephdecay-teff}) on ephemeris deterioration. The rate of ephemeris deterioration seems to depend on the planet radius. This is easily explained for the larger planets: SNR generally increases with planet size, and $\delta T_{0,\tess}$ (and thus $\delta P$) is inversely proportional to the SNR. However, this trend changes direction around 5 $R_{Earth}$, and the rate of ephemeris deterioration decreases with size below this $R_p$ value. We believe this effect is due to a correlation between the radii and periods of the simulated planets. Indeed, we find that below this $R_p$ threshold, the fraction of simulated planets with $P < 10$ days vs. $P > 10$ days {\it increases} with decreasing R$_p$. However, the fact that smaller planets are harder to detect at longer periods with \tess\ likely contributes to this effect as well.

The fraction of planets with $\delta T_{0,m} < 30$ min does not significantly depend on either the \tess\, magnitude or the effective temperature of the host stars. Some large changes in this ratio are apparent for some values of these two parameters, but these fluctuations correspond to bins with very small number statistics and are thus unlikely to be significant.

\section{Ephemeris Expiration Analysis and Results Using Real \tess\ Planets and Planet Candidates}
\label{sec:toianalysisresults}

Since over 1000 real planet candidates (known as \tess\ Objects of Interest, or TOIs) have already been released, we use equation \ref{eqn:delTc} to also determine $\delta T_{0,1y}$ for TOIs in order to cross-check our results based on the simulated planets. We use the $\delta P$ and $\delta T_{0,TESS}$ values provided by the \tess\ Science Office (TSO) for each TOI (available on ExoFOP-TESS \footnote{https://exofop.ipac.caltech.edu/}).

\begin{figure*}
    \centering
    \includegraphics[scale=0.33]{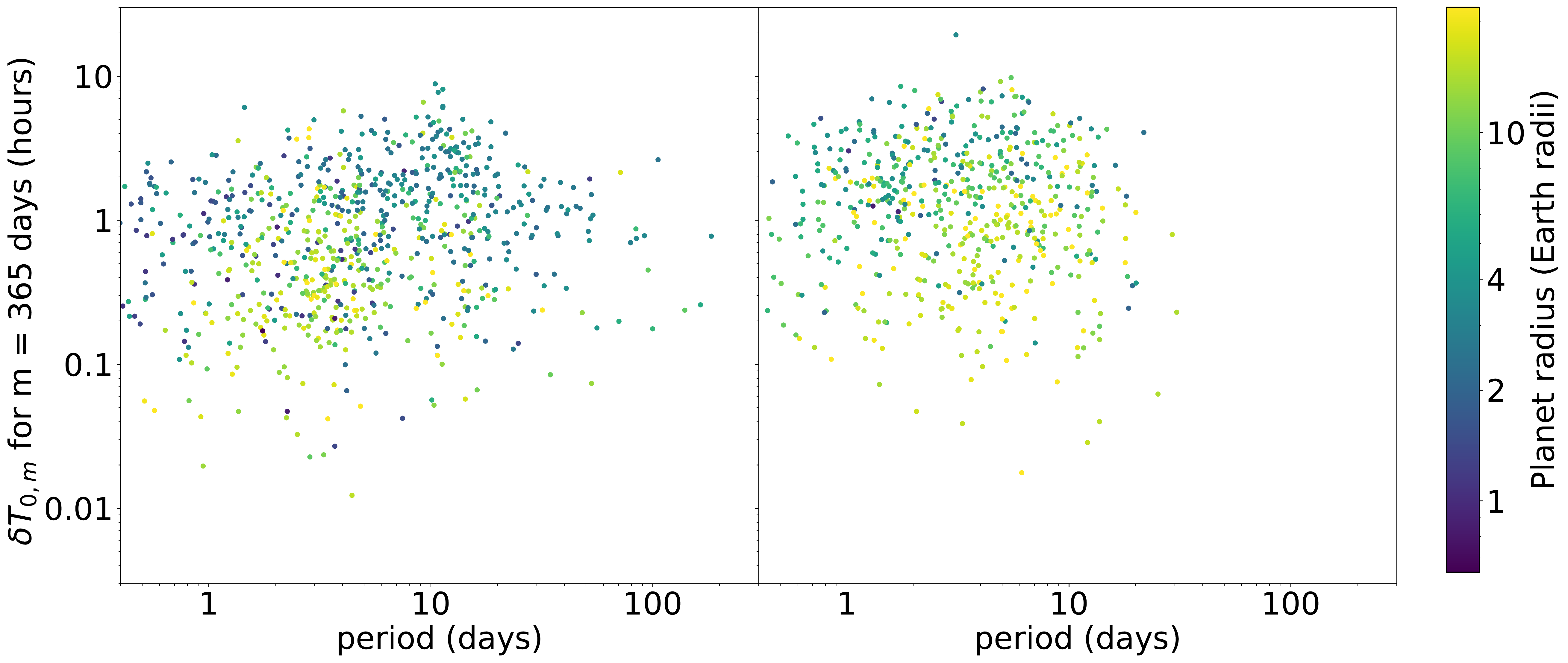}
    \caption{The uncertainty in mid-transit time for real TOIs one year after \tess\, observes them, as a function of orbital period. The colors represent planet radius. {\it Left:} TS planets. {\it Right:} FFI planets. }
    \label{fig:ephdecaytois-summary}
\end{figure*}

Figure \ref{fig:ephdecaytois-summary} shows the 1-year ephemeris deterioration for the 1604 TOIs available on Jan. 15, 2020, and can be compared to Figure \ref{fig:ephdecay-summary}. We caution that the TOI sample contains an unknown number of false positives (primarily eclipsing binaries), and is biased towards larger and shorter-period planet candidates (which are easier to detect). Despite these biases, we find similar features between the simulated planet sample and the TOI sample: the ephemerides of the TS TOIs deteriorate slower than those of the FFI TOIs; larger planets generally have smaller $\delta T_{0,1y}$; and there is a peak in the distribution just past $P\sim10$ days (this feature is not as obvious in the FFI TOI sample, likely because the period space hasn't been sufficiently populated at longer periods). We thus feel confident in the accuracy of the predictions, interpretation and recommendations for ephemeris maintenance that we present in this paper based on the simulated planet sample.

\section{Discussion}
\label{sec:discussion}

\subsection{Considerations for \jwst\, observations}

Transmission spectroscopy with \jwst\, represents the most widely anticipated type of TCO, and we do not expect nor recommend that \jwst\, will observe transits with transit mid-point uncertainty greater than 30 min, particularly if this uncertainty can be reduced by additional ground-based observations. In this section we examine the ephemeris deterioration of \tess\, planets as a function of their suitability for \jwst\, observations.

\begin{figure*}[ht]
    \centering
    \includegraphics[scale=0.3]{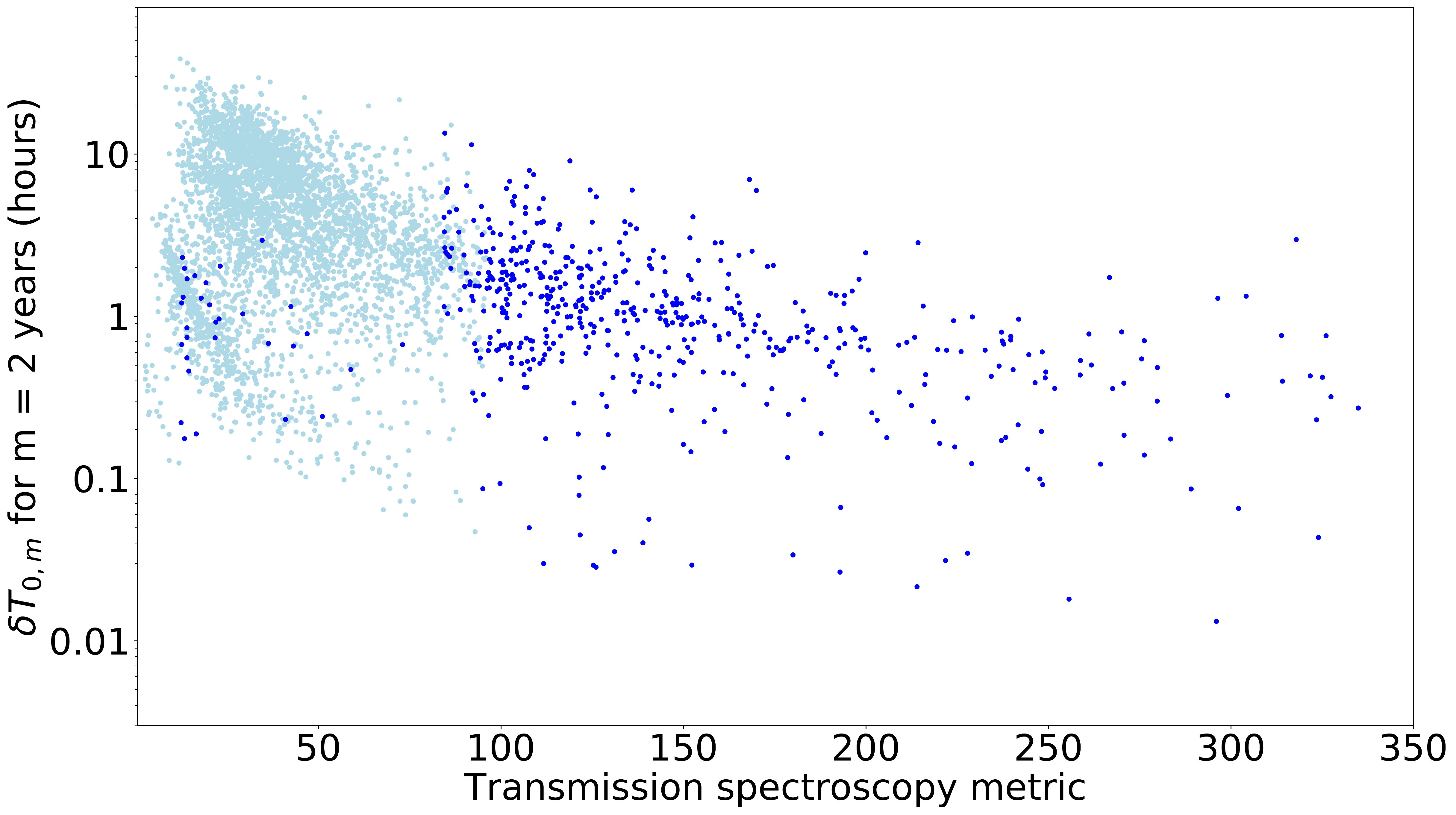}
    \caption{The uncertainty in mid-transit time for simulated planets two years after \tess\, observes them, as a function of the transmission spectroscopy metric. Planets above the TSM thresholds listed in Table 1 of \cite{Kem18} are highlighted in dark blue.}
    \label{fig:eph-TSM}
\end{figure*}

We use $m = 2$ years as a representative value for the average timespan between the $T_{0,\tess}$ of a typical PM planet (i.e. observed in July 2019), and the time when \jwst\, should begin science operations (i.e. six months after its currently planned launch date of early 2021). In Figure \ref{fig:eph-TSM} we show $\delta T_{0,m}$ evaluated two years after the end of a planet's \tess\, observations as a function of the transmission spectroscopy metric (TSM) described in \cite{Kem18}. Briefly, the TSM corresponds approximately to the S/N for 10 hours (with 5 hours occurring during transit) of observations with the NIRISS instrument on \jwst, under the assumptions made in \cite{Kem18}. Table 1 of \cite{Kem18} lists cutoff TSM values corresponding to the top \tess\, planets for atmospheric characterization. In Figure \ref{fig:eph-TSM} we highlight in dark blue the 395 planets that are above those cutoff values.

The TSM correlates with the \tess\, SNR, so in general the planets having the highest TSM experience less ephemeris deterioration than those with lower TSM values. However, under our assumptions, the majority of the planets (313 out of 395, or 79\%) recommended by \cite{Kem18} will still have $\delta T_{0,m} > 30$ min by the time they would be observed with \jwst\, (if no follow-up transits are observed). In particular, for small ($R_p < 4 R_{Earth}$) planets with TSM above the \cite{Kem18} thresholds, the ephemerides of 64 out of 81 transiting M dwarfs, and 56 out of 60 transiting F/G/K dwarfs, will have expired by the time \jwst\ begins science observations. Most \jwst\ targets will thus require ephemeris ``refreshment" (i.e. reducing $\delta T_{0,m}$ to less than 30 min.) prior to scheduling them for observations (unless they have been re-observed as part of the extended \tess\ mission before the \jwst\ scheduling takes place).

\subsection{Resources for keeping ephemerides fresh}

Transits deeper than $\sim$1000 ppm (e.g. \citealt{Gue19a}) and with durations shorter than $\sim$ 7 hours can generally be recovered with ground-based meter-class telescopes. The \tess\, Follow-Up Observing Program (TFOP) subgroup 1b (SG1b) focuses on ground-based photometric follow-up. SG1b marshals tens of telescopes for follow-up photometry to verify \tess\, Objects of Interest (TOIs) to either confirm them as planets or identify false positives (FPs). In so doing, these efforts  also refresh the ephemerides for those planets whose transits they observe.

To keep fresh ($\delta T_{0,m} <$ 30 min) the ephemerides of planets with long or shallow transits, which make up 69\% of all the simulated planets, it will be necessary to use space-based observatories. TFOP SG5 coordinates a number of space-based follow-up efforts toward this goal. There are three recent or current space-based observatories that can realistically be used for this purpose: \cheops, \spitzer, and \hst. We also consider the impact of the extended \tess\, mission in the next subsection (\ref{sec:ext}).

\subsubsection{\spitzer}
The now defunct \spitzer\, space telescope has a 85-cm aperture. In its warm phase, it could observe in two channels, 3.6 and 4.5 $\mu$m. Thanks to its Earth-trailing orbit, \spitzer\, could observe any target for at least $\sim$80 days per year. \spitzer\, has already rescued the ephemerides of several \ktwo\ planets (e.g. \citealt{Ben17, Liv19, Kos19}). A recent large program to achieve the same goal for the \tess\, planets most amenable to atmospheric characterization has improved the ephemerides of 34 TOIs (I. Crossfield, private communication) before the \spitzer\, mission was terminated in January 2020. While valiant, this effort only scratches the surface of the more than 2000 \tess\ planets that are expected to require space-based follow-up in order to refresh their ephemerides (see section \ref{sec:recs}).

\subsubsection{\hst}

\hst\, has a 2.4-m aperture, and can observe \tess\, transits with much higher SNR than \tess. \hst\, has an equatorial orbit, part of which it spends between the Earth and the Sun, so most of the sky cannot be observed continuously and many transit observations will not sample the full transit. The transit time precision of HST should still be sufficient for ephemeris refreshment.\footnote{Even for the 1000 ppm transit of HD 97658b, the uncertainty on $T_0$ is only 8 minutes \citep{Knu14b}, which is sufficient for long-term ephemeris refreshment as long as the time elapsed between the end of the \tess\, observations and the \hst\ transit observation is long enough (see section \ref{sec:recs}).} We expect that a number of \tess\, planets will be proposed for atmospheric characterization, particularly in the years prior to \jwst. \tess\, planets with very shallow transits may even be proposed solely for ephemeris refinement, especially since \spitzer\ is no longer available. Assuming the corresponding \hst\, observations themselves are scheduled before $\delta T_{0,m}$ becomes too large, a lucky few \tess\, planets will have their ephemerides refreshed during transmission spectroscopy observations.

\subsubsection{\cheops}
The European Space Agency successfully launched the CHaracterising ExOPlanets Satellite (\cheops; \citealt{Broeg13}) on December 18, 2019, and the satellite is currently in its commissioning phase. \cheops\, has a 30-cm aperture and its passband spans the 0.4 - 1 $\mu$m range. Only 20\% of \cheops\, observing time is open and allocated through an ESA Guest Observer program, but the \cheops\, consortium may observe transits of \tess\, planets as part of the Guaranteed Time Observing program (which manages the remaining 80\% of \cheops\, time). It is anticipated to achieve significantly better photometric precision than \tess\, thanks to its larger aperture. There are two downsides of \cheops\, that are important to recognize for its role in ephemeris refreshment. The first is that large portions of the \tess\, footprint surrounding the ecliptic poles (where \tess\, is expected to discover a disproportionate number of planets) will not be observable by \cheops\, due to the operational and pointing constraints of its orbit. The second is that it is in low Earth orbit and for most stars observations will be periodically interrupted by the Earth. However, as for \hst\, observations, the transit time precision should still be amply sufficient for ephemeris refreshment.

\subsection{Extended \tess\, mission}
\label{sec:ext}

Perhaps the most compelling resource for preventing deteriorated ephemerides is the extended \tess\, mission. This 2.5-year extension (including a repeat of year 1 in year 3, and a partial repeat of year 2 in year 4) to the \tess\, PM was approved in summer of 2019. Thus, a large fraction of planets will be re-observed by \tess\, approximately two years after their PM observations. However, as shown in Figure \ref{fig:ephdecay-summary}, the ephemerides of most \tess\, planets will have expired just one year after their \tess\, observations, making it difficult to schedule TCOs such as transit spectroscopy with the \hst\, or ground-based facilities, or Rossiter-McLaughlin observations, in the 1-2 years until \tess\ re-observes these targets. Moreover, the extended mission will not have refreshed the ephemerides of many \tess\, planets (particularly those from the northern ecliptic hemisphere) in time for Cycle 1 of \jwst\, making it challenging or even impossible to schedule observations unless follow-up transits are observed with other facilities.

\begin{figure*}
    \centering
    \includegraphics[scale=0.33]{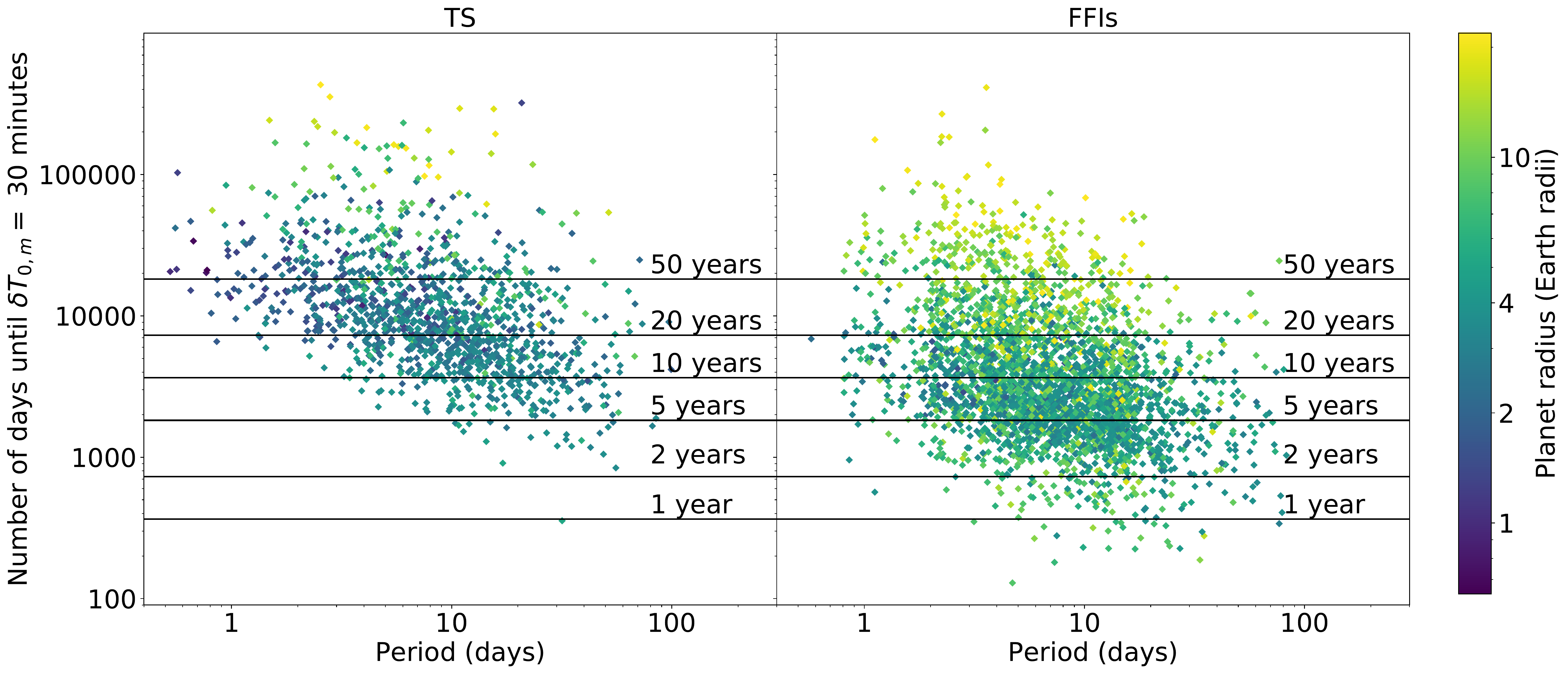}
    \caption{Number of days ephemerides stay fresh ($N_{\delta T_{0,m} < 30 {\rm min}}$), starting from 2-minute cadence extended mission \tess\ observations assumed to take place two years after the PM observations.}
    \label{fig:ext2min}
\end{figure*}

\begin{figure*}
    \centering
    \includegraphics[scale=0.33]{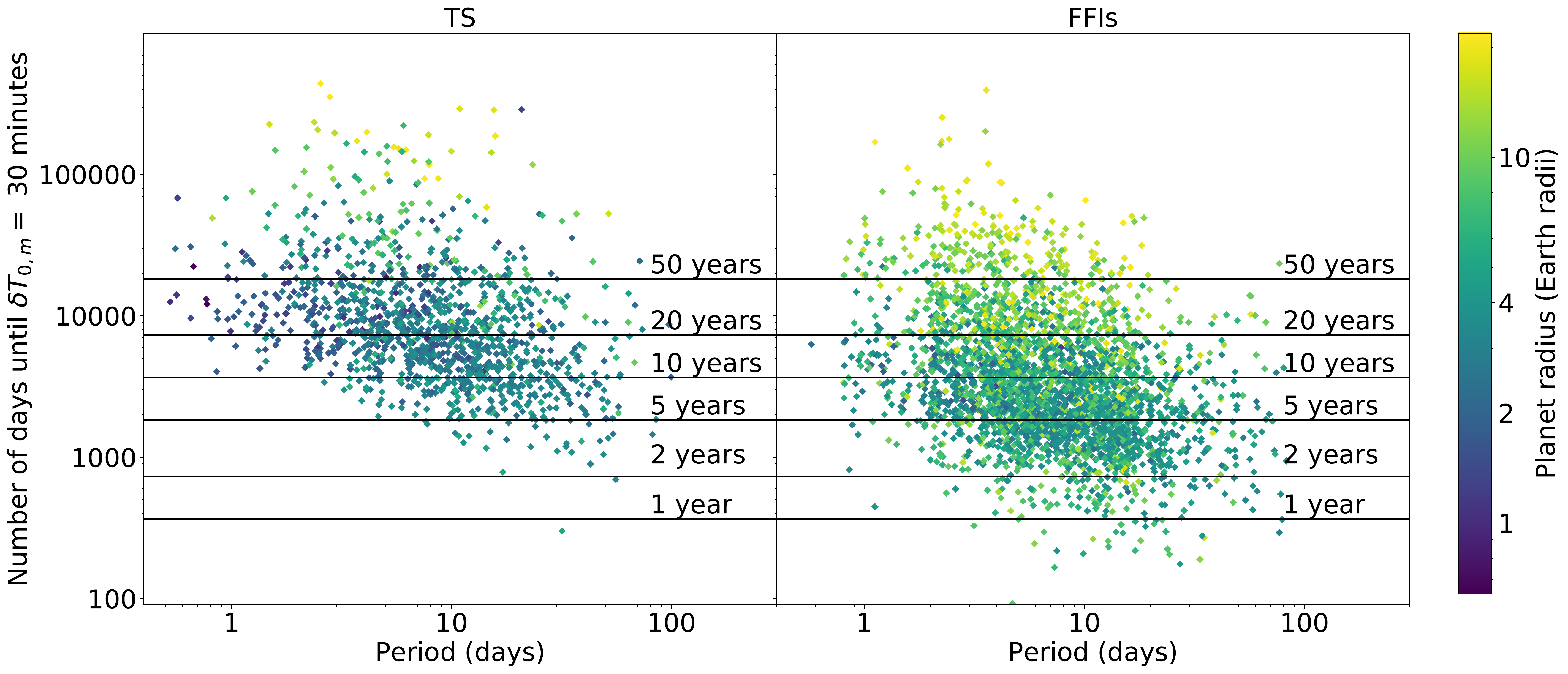}
    \caption{Same as Figure \ref{fig:ext2min}, but assuming a 10-minute cadence for the extended mission observations.}
    \label{fig:ext10min}
\end{figure*}

Even for TCOs that will take place during or after the extended mission, there are a few important caveats. Approximately 5-10\% of PM \tess\ planets and candidates will fall in the gaps between sectors during the extended mission, and will not be re-observed. In addition, as currently planned, \tess' year 4 will only include $\sim$65\% of the PM northern ecliptic hemisphere, leaving many northern TOIs un-observed. Lastly, the extended mission will find hundreds to thousands of new planets \citep{Hua18, Bou17} whose ephemerides will eventually need to be rescued as well. Therefore, we recommend establishing an ephemeris refreshment procedure for \tess\, planets and planet candidates to address these caveats and to complement the extended mission.

Nevertheless, since the majority of PM planets and candidates will be re-observed during the extended mission, we investigate how long their ephemerides will remain fresh after their second round of observations. During the extended mission, \tess\, will again observe $\sim$200 000 targets/year at 2-minute cadence, while full frame images will be taken every 10 minutes (instead of 30 minutes as was done during the PM). Since at least 80\% of the short cadence targets will be selected through the Guest Investigator program from among targets proposed by the community, we do not yet know the properties of those targets. Therefore, we perform our investigation using our simulated yield planets for both cadences. Figure \ref{fig:ext2min} shows the length of time for which the PM \tess\, planet ephemerides remain fresh ($N_{\delta T_{0,m} < 30 {\rm min}}$), from the time of hypothetical 2-minute cadence extended mission observations (assumed to take place two years after the PM observations). Figure \ref{fig:ext10min} is identical except that we assume 10-minute cadence during the extended mission.

These figures primarily show that even with the extended mission, the ephemerides of most PM \tess\ planets will expire in as little as 10 years, unless they are refreshed again (either through another \tess\ extension, or transit observations with other telescopes). This 10-year timescale is comparable to timelines for major post-\jwst\ facilities like ARIEL, GMT, E-ELT and TMT, as well as more distant, prospective observatories (e.g. LUVOIR, Origins Space Telescope, etc.). 

By eye, there are no large differences between Figures \ref{fig:ext2min} and \ref{fig:ext10min}. Indeed, $N_{\delta T_{0,m} < 30 {\rm min}}$ varies by less than 50\%, comparable to the likely uncertainty in the first light time for the facilities listed above.

\subsection{Recommendations for keeping \tess\, ephemerides fresh}
\label{sec:recs}

\begin{figure*}
    \centering
    \includegraphics[scale=0.3]{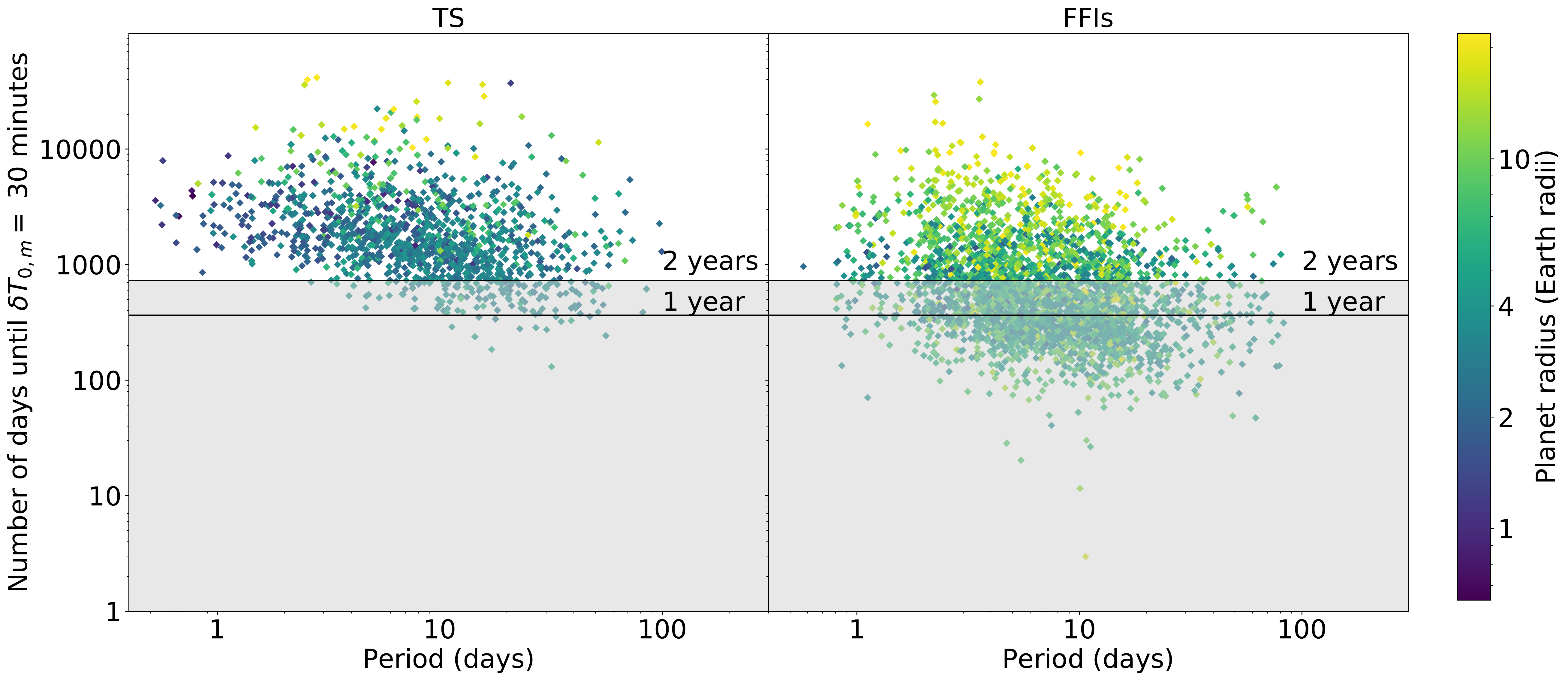}
    \includegraphics[scale=0.3]{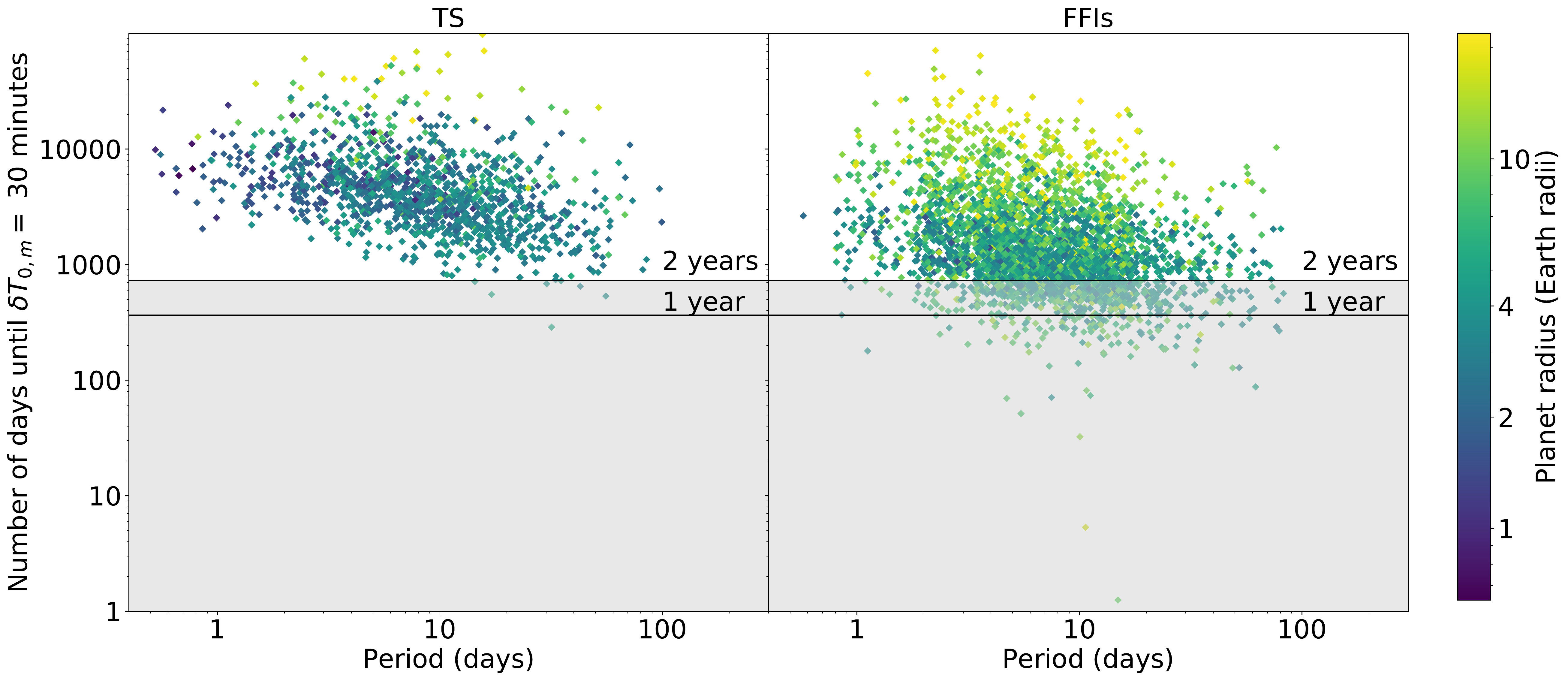}
    \caption{{\it Top:} Number of days ephemerides stay fresh ($N_{\delta T_{0,m} < 30 {\rm min}}$) starting from a follow-up transit observed three months after the end of each planet's \tess\, observations. The shaded area represents $N_{\delta T_{0,m} < 30 {\rm min}}$ $<$ 2 years. {\it Bottom:} Same as top, but after a follow-up transit observed nine months after the end of each planet's \tess\, observations.}
    \label{fig:eph-refresh}
\end{figure*}

We investigated the impact of the follow-up baseline (i.e. the time elapsed between the end of the \tess\, observations and the follow-up transit observation), as well as the SNR and the cadence of the follow-up observations. For a fixed SNR and baseline, the choice of cadence of the follow-up observations only changes the length of time that the ephemeris stays fresh by at most a few percent, if ingress and egress are well-sampled. However, if the cadence is such that fewer than one observation is taken during ingress or egress (see equation \ref{eqn:sigtc2}), then the amount of time the ephemeris stays fresh (i.e. $\delta T_{0,m} < 30$ min) can change by tens of percent, compared to a transit with several observations during ingress and/or egress. We note that the planets most at risk (periods longer than of order 10 days) are also those with longer ingress and egress durations, for which the cadence of follow-up observations matters least, as long as the sampling rate is not longer than a few minutes. The SNR of the follow-up observations is more important since it is inversely proportional to the uncertainty on the mid-transit time of the follow-up transit ($\delta T_{c,m}$). 

The following equation shows how $\delta P$ depends on $\delta T_{0,\tess}$, $\delta T_{c,m}$ and the number of orbital cycles elapsed between the two ($n_m$):
\begin{equation}
\label{eqn:delP}
\delta P = \sqrt{\frac{\delta T_{0,\tess}^2 + \delta T_{c,m}^2}{n_{m}}}
\end{equation}

We see from equation \ref{eqn:delP} that while $\delta P$ decreases with decreasing $\delta T_{c,m}$, it decreases faster with increasing $n_{m}$. Therefore, the most important variable to consider when planning follow-up transit observations is the baseline.  In essence, a follow-up transit should be obtained as long as possible after the \tess\, observations (but while $\delta T_{0,m}$ is still small enough to allow for scheduling the follow-up observations).

Based on these considerations, we present an ephemeris refreshment plan that can reliably refresh the ephemerides of the vast majority of \tess\, planets for at least two years from their corresponding $T_{0,\tess}$, with just one transit per planet. We conservatively assume that transits deeper than 1000 ppm and with durations shorter than 7 hours (610 TS and 1542 FFI planets) can routinely be followed up from the ground, so for every simulated planet with a transit depth above this threshold, we calculated the SNR achievable with a 1.0m telescope in I band. We added in quadrature shot noise, scintillation noise (using equation 1 of \citealt{Man11}) and atmospheric noise (estimated at 400 ppm, following \citealt{Man11}) to estimate the total photometric noise. We assumed average airmass (1.3), as well as an exposure time and overhead of 30 s each (typical of ground-based observations with meter-class telescopes), for an overall sampling rate of 60 s. For each planet with transits shallower than 1000 ppm or longer than 7 hours (686 TS and 1538 FFI planets), we estimated the SNR that would be reached with \spitzer\, at 4.5 $\mu$m, assuming an exposure time of 2 s and negligible overhead, which is typical of the majority of \spitzer\, exoplanet observations \footnote{As described above, the cadence of the observations has minimal impact on the effectiveness of a follow-up transit for ephemeris refreshment.}. 

We used equation \ref{eqn:delP} to estimate $\delta P$ after the addition of a follow-up transit observation, and equation \ref{eqn:delTc} (replacing $\delta T_{0,\tess}$ with $\delta T_{c,m}$) with $\delta T_{0,m}$ set to 30 min. to determine the improvement in the ephemerides after these follow-up observations. We examined how long it would take until the renewed ephemerides deteriorate again. Figure \ref{fig:eph-refresh} shows the length of time for which the \tess\, planet ephemerides remain fresh ($N_{\delta T_{0,m} < 30 {\rm min}}$) with just one follow-up transit observed three (top) and nine (bottom) months after the end of the \tess\, observations of each planet. 

We find that the ephemerides of 89\% of the TS planets and only 38\% of the FFI planets can be refreshed for at least two years from the follow-up transit observed at three months. In the context of \jwst\, observations, this strategy should be sufficient for scheduling almost any of the northern ecliptic hemisphere TS planets \tess\, finds, since the \jwst\, Cycle 1 observations are expected to happen approximately two years from the second half of the \tess\, PM survey. For the remaining northern hemisphere planets (including most of the FFI planets), and for many of those in the southern ecliptic hemisphere, a longer baseline between the \tess\, observations and the follow-up transit (e.g. nine months) will be necessary to sufficiently refresh their ephemerides, if those planets are to be observed during cycle 1 of \jwst.

We also examine the improvement in ephemeris refreshment with a transit observation nine months after the end of \tess\, observations. Figure \ref{fig:eph-refresh} shows that the longer baseline refreshes the ephemerides of 99\% of all TS planets, and the vast majority (80\%) of FFI planets. While the longer nine-month baseline is more effective, in many cases the initial \tess\ ephemeris would have already expired by $m = 9$ months. For those planets, transit follow-up observations should ideally be done both three {\it and} nine months from $T_{0,\tess}$.

Since our analysis does not account for TTVs, we recommend that for any system suspected of harboring more than one planet, observers should obtain an estimate of the amplitude of possible TTVs, and consider it when scheduling follow-up transit observations. However, we note that only $\sim$20 systems that \tess\, will find are expected to show measurable TTVs \citep{Had19}, so we do not expect this to be a consideration for preserving the ephemerides of the majority of \tess\, planets.

Finally, while we expect that observers interested in individual \tess\, systems will take the initiative to ensure their ephemerides are refreshed prior to scheduling \jwst\ (via TFOP SG1, SG5, or otherwise), or other expensive observations, we also summarize here the categories of planets most at risk of ephemeris deterioration for observers wishing to refresh \tess\, ephemerides in bulk:

\begin{itemize}
\item Planets that will not be re-observed during the extended mission;
\item FFI planets in general (whose ephemerides will deteriorate faster than those of TS planets);
\item TS planets with $4 \lessapprox P \lessapprox 40$ days (whose ephemerides become uncertain faster than for planets with shorter or longer periods);
\item TS planets with $R_p \lessapprox 5 R_{Earth}$. 
\end{itemize}

\subsection{False positive rate considerations}

Ideally, observations for ephemeris refreshment (particularly those that require space-based or larger ground-based telescopes) would only be carried out for confirmed \tess\, planets. \cite{Sul15} and \cite{Bar18} estimated \tess\, false positive rates for TSs ($\sim$50\%) and FFIs ($\gtrsim$85\%), respectively. While it is still too early in the mission to know the true rates, the false positive rate will be higher for FFI candidates (with the planets coming from this sample also being in the most dire need of ephemeris refreshment). However, standard vetting of TOIs (odd/even eclipse tests, centroid analyses, visual inspection, etc.) is already identifying a large number of false positives. TFOP efforts are separating false positives from planets efficiently, within a few weeks for the most interesting TOIs. 

By the time TOIs go through basic TFOP observations (to be confirmed as planets or ruled out as false positives), assuming this would happen two months from $\delta T_{0,\tess}$, \footnote{Even though this has been the case for numerous TOIs (given that data releases are happening within a few weeks of the end of a sector), we note that the sheer number of TOIs is such that TFOP resources may not be sufficient to follow-up all of them before they set for the season.}, we expect that 209 (141 from FFIs and 68 from TSs) of the current (i.e. 1604) TOIs have $\delta T_{0,2months}> 30$ min. Approximately 92 of those should be detectable from the ground and their ephemerides will be (temporarily) refreshed as part of the seeing-limited photometry step (under the umbrella of TFOP SG1b), {\it if it can happen within 2 months of their \tess\ observations}. For the remaining 117, space-based photometry would be urgently required, before their ephemerides deteriorate further.

Note that Figure \ref{fig:ephdecaytois-summary} may be of interest for TFOP photometric follow-up efforts of TOIs that will turn out to be FPs as well, since the figure (particularly the larger planet candidates, which are more likely to be EBs) very likely contains a number of FPs. 

\section{Conclusion}
\label{sec:conclusion}

Ephemeris deterioration constitutes a major problem that can impede exoplanet follow-up observations that need to be acquired at a specific time of the planet's orbit (most frequently during transit). While the efforts of programs such as the Transit Ephemeris Refinement and Monitoring Survey (TERMS; \citealt{Kan09}) have successfully refreshed the ephemerides of nearly a dozen transiting exoplanets (e.g. \citealt{Drag11}), the ephemerides of many known transiting exoplanets have been thoroughly lost. These include the ephemerides of most CoRoT planets and planet candidates (H. Deeg, private communication; \citealt{Dee15}), as only a handful have been re-observed since their CoRoT observations were taken over five years ago \citep{Rae19}. A similar fate likely awaits \kepler\ planets (with a few exceptions such as Kepler-167e; \citealt{Dalba19}) and even many \ktwo\ planets if measures are not taken to maintain their ephemerides fresh.

In this work, we investigated the extent and progress of ephemeris deterioration for a simulated yield of \tess\, planets, and cross-checked our results using the current sample of real \tess\ planets and planet candidates (i.e. TOIs). We studied the ephemeris expiration timescale as a function of several planetary and stellar parameters, for both 2-minute and 30-minute cadence planets. We found that the ephemerides of 81\% of simulated \tess\ planets will be expired one year after their \tess\ observations. We found that the ephemerides of the planets observed with the longer cadence become uncertain faster due to the lower precision on the transit times, which in turn leads to a lower $\delta P$ measured from the \tess\, light curves. We also found that the ephemerides of planets with short or long periods deteriorate slower than those of planets with $4 \lessapprox P \lessapprox 40$ days. 

The approval of a 2.5-year extension for \tess\ has significantly improved ephemeris refreshment prospects for PM planets. We find that the ephemerides of most PM planets will be refreshed for at least 10 years past their extended mission observations. We also find that whether these new observations are taken at a 2-minute versus 10-minute cadence does not impact this timescale nearly as much as the fact that they are taken two years after the original observations, in line with our section \ref{sec:recs} finding that the cadence has a relatively low impact on the effectiveness of a new transit for ephemeris refreshment, compared to the baseline.

However, due to the timeline and observing strategy of the extended mission, follow-up transit observations will still be necessary for TCOs intended to take place over the next 1-2 years, in order to prevent ephemeris deterioration. Moreover, a number of PM planets will not be re-observed during the extended mission, so those planets should be prioritized for follow-up ephemeris refreshment observations. For sufficiently deep and short transits, this can be achieved with the multitude of ground-based telescopes that participate in TFOP SG1 activities. Critically, for shallower or longer transits (which make up half of the simulated planets), space-based telescopes such as \hst\, or \cheops\, are needed. The longer the baseline between the \tess\ and the follow-up observations, the longer the ephemerides will stay fresh. We find that for 98\% of expected \tess\, planets, one or two follow-up transits observed three and/or nine months after the end of a planet's \tess\, observations will refresh its ephemeris for two years past the follow-up observations.

The ephemeris refreshment strategy we describe in this paper and the \tess\ extended mission should be sufficient for scheduling TCOs for \tess\ planets for the next few to several years. However, more distant TCOs (even for planets re-observed by \tess, but also for new planets discovered during the extended mission) with the future Extremely Large Telescopes and missions such as Ariel \citep{Ecc16} (and beyond) will still eventually require additional transit follow-up to maintain fresh ephemerides. Alternatively, additional extensions to the \tess\ mission will solve much of this problem, and preserve the ephemerides of \tess\ planets ready for characterization for decades to come.

\section{Acknowledgments}

This paper includes data collected by the \tess\ mission, which are publicly available from the Mikulski Archive for Space Telescopes (MAST). Funding for the \tess\ mission is provided by NASA's Science Mission directorate. We acknowledge the use of public \tess\ Alert data from pipelines at the \tess\ Science Office and at the \tess\ Science Processing Operations Center. Resources supporting this work were provided by the NASA High-End Computing (HEC) Program through the NASA Advanced Supercomputing (NAS) Division at Ames Research Center for the production of the SPOC data products. JP acknowledges funding support from the NSF REU program under grant number PHY-1359195.

We thank the anonymous referee for feedback which has significantly improved the clarity of the paper. The authors thank Karen Collins, Chelsea Huang, Robert Zellem, Stephen Kane, Luke Bouma, Laura Kreidberg, Sam Quinn, Sam Hadden and Jacob Bean for suggestions of figures and discussion points to make the paper useful to a wide range of observing interests. DD acknowledges support provided by NASA through Hubble Fellowship grant HSTHF2-51372.001-A awarded by the Space Telescope Science Institute, which is operated by the Association of Universities for Research in Astronomy, Inc., for NASA, under contract NAS5-26555. TB acknowledges support from the Sellers Exoplanet Environments Collaboration.

\bibliographystyle{aasjournal}
\bibliography{research}

\end{document}